\documentclass[pra,aps,twocolumn,a4paper,showpacs,superscriptaddress,10pt]{revtex4}

\usepackage{amsmath}
\usepackage{amssymb}
\usepackage{graphicx}
\usepackage{color}
\usepackage{hyperref}
\usepackage[USenglish]{babel}
\newcommand{\mytitle}{Controlling the $2p$ Hole Alignment in Neon via the $2s$--$3p$ Fano Resonance}

\newcommand{\rmpdfinfo}{\special{ps:: userdict /pdfmark /cleartomark load put}}

\definecolor{MyDarkGreen}{rgb}{0,0.6,0}
\definecolor{MyDarkBlue}{rgb}{0,0,0.8}
\definecolor{MyDarkRed}{rgb}{0.6,0,0.3}
\hypersetup{breaklinks=true, colorlinks=true,plainpages=true, linktocpage=true, linkcolor=MyDarkBlue, citecolor=MyDarkGreen, urlcolor=MyDarkRed, pdfborder={0 0 0},%
pdfauthor={Elisabeth Heinrich-Josties, Stefan Pabst, and Robin Santra},%
pdfsubject={Research article},
pdftitle={\mytitle}%
}

\newcommand{\figwidth}{.9}
\newcommand{\sket}[1]{\left|#1\right)}
\newcommand{\sbra}[1]{\left(#1\right|}
\newcommand{\ket}[1]{\left|#1\right>}

\newcommand{\half}[0]{\frac{1}{2}}

\begin{document}

\title{\mytitle} 

%\author{Elisabeth Heinrich-Josties$^{1,2}$, Stefan Pabst$^1$, and Robin Santra$^{1,3}$}
%\address{$^1$ Center for Free-Electron Laser Science, DESY, Notkestrasse 85, 22607 Hamburg, Germany}
%\address{$^2$ University of California Los Angeles, Department of Physics \& Astronomy, CA, USA}
%\address{$^3$ Department of Physics, University of Hamburg, Jungiusstrasse 9, 20355 Hamburg, Germany}

\author{Elisabeth Heinrich-Josties}
\affiliation{Center for Free-Electron Laser Science, DESY, Notkestrasse 85, 22607 Hamburg, Germany}
\affiliation{University of California Los Angeles, Department of Physics \& Astronomy, CA, USA}
%\email[]{}

\author{Stefan Pabst}
%\email[]{stefan.pabst@cfel.de}
\affiliation{Center for Free-Electron Laser Science, DESY, Notkestrasse 85, 22607 Hamburg, Germany}

\author{Robin Santra}
%\thanks{Corresponding author}
%\email[]{robin.santra@cfel.de}
\affiliation{Center for Free-Electron Laser Science, DESY, Notkestrasse 85, 22607 Hamburg, Germany}
\affiliation{Department of Physics, University of Hamburg, Jungiusstrasse 9, 20355 Hamburg, Germany}

\date{\today}

%%  abstract  %%
%%%%%%%%%%%%%%%%

\begin{abstract}
We study the state-resolved production of neon ion after resonant photoionization of Ne via the $2s$--$3p$ Fano resonance. 
We find that by tuning the photon energy across the Fano resonance a surprisingly high control over the alignment of the final $2p$ hole along the polarization direction can be achieved.
In this way hole alignments can be created that are otherwise very hard to achieve.
%Without the electron-electron correlation responsible for the autoionization of the $2s^{-1}3p$ state, the variation in the hole alignment disappears.
The mechanism responsible for this hole alignment is the destructive interference of the direct and indirect (via the autoionizing $2s^{-1}3p$ state) ionization pathways of $2p$. By changing the photon energy the strength of the interference varies and $2p$-hole alignments with ratios up to 19:1 between $2p_0$ and $2p_{\pm 1}$ holes can be created---an effect normally only encountered in tunnel ionization using strong-field IR pulses.
%The destructive interference of the direct $2p$ ionization pathway and the indirect $2p$ ionization pathway (via the autoionizing $2s^{-1}3p$ state) creates a dark state in the continuum leading to a suppression in the overall $2p$ ionization.
%Particularly strong is the suppression in the $2p^{-1}\varepsilon d$ partial-wave channel resulting in $2p$-hole alignment with ratios up to 19:1 between $2p_0$ and $2p_{\pm 1}$ holes---an effect normally only encountered in tunnel ionization using strong-field IR pulses.
Including spin-orbit interaction does not change the qualitative feature and leads only to a reduction in the alignment by $2/3$. %of the hole alignment but reduces the maximum achievable hole alignment by $2/3$.
Our study is based on a time-dependent configuration-interaction singles (TDCIS) approach which solves the multichannel time-dependent Schr\"odinger equation.
\end{abstract}

% 31.15.A-  Ab initio calculations 
% 32.80.Rm  Multiphoton ionization and excitation to highly excited states 
% 32.30.Jc  Visible and ultraviolet spectra
% 32.80.Zb  Autoionization
% 42.65.Re  Ultrafast processes; optical pulse generation and pulse compression
\maketitle

%%  main part  %%
%%%%%%%%%%%%%%%%%

%intro: more focus on control less on high intensity
%
%results in few fields:
% - add hole alignment discussion including explaining origin
%
%results in intense fields:
%- first ionization dynamics (for two intensities show the three frequency cases for all shells)
%- hole alignment in 
%- final hole populations: stable 2s hole (and signature of rabi)

\section{Introduction}
\label{sec1}
%------------------------------------------------------------

%Fano
Fano resonances~\cite{Fa-PhysRev-1961} appear in almost any field of physics ranging from atomic physics to solid-state physics and to optics~\cite{MiFl-RMP-2010}.
Their most characteristic feature is the asymmetric line profile~\cite{KrPa-AJP-2014} which results from the coherent interference of a direct continuum channel and an indirect channel which involves a discrete quasi-bound state~\cite{BaDu-AJP-2004}.
These asymmetric line shapes have been first discussed in atomic physics in the context of photoabsorption~\cite{Be-ZPhys-1935} and electron scattering~\cite{La-RadResSupp-1959}.

%%%% use Fano resonance and manipulate them
%%%%%%%%%%%%%%%%%%%%%%%%%%%%%%%%%%%%%%%%%%
In the last decades, there has been an increasing interest in Fano resonances in the presence of strong-field~\cite{LaZo-PRA-1981,TaGr-PRA-2012} and ultrashort~\cite{WiBu-PRL-2005,WaCh-PRL-2010} pulses.
%The interest in Fano resonances and autoionizing states in atomic systems has been increasing again with the advent of strong-field~\cite{LaZo-PRA-1981,TaGr-PRA-2012} and ultrashort pulses~\cite{WiBu-PRL-2005,WaCh-PRL-2010}.
Strong-field pulses modify the ionization continuum and alter~\cite{OtKa-Science-2013} or even destroy~\cite{TaGr-PRA-2012} the characteristic Fano profile.
With attosecond and femtosecond pulses the electron motion of an autoionization process can be studied~\cite{ArLi-PRL-2010,GiCh-RPL-2010}.
%A popular technique to probe the dynamics of the autoionizing states, when exposed to a short, intense laser pulse, is transient absorption spectroscopy~\cite{LoGr-ChemPhys-2007,WaCh-PRL-2010,ChBe-PRA-2012,OtKa-Science-2013}.
Also the interplay of Fano resonances with free-electron laser pulses has been investigated~\cite{NiMe-PRA-2009}.

% Phys. Rev. A 80, 055402 (2009) [4 pages]
%Time-dependent formation of the profile of resonance atomic states and its dependence on the duration of ultrashort pulses from free-electron lasers
% Phys. Rev. Lett. 105, 053002 (2010) [4 pages]
%Ionization Branching Ratio Control with a Resonance Attosecond Clock
% Phys. Rev. Lett. 105, 263003 (2010) [4 pages]
%Monitoring and Controlling the Electron Dynamics in Helium with Isolated Attosecond Pulses

%hole alignment (control)
With the rapid advances in laser technology, it is nowadays possible to study the influence of the details of the ionization process on the parent ion and not only on the ionized photoelectron~\cite{LoLe-PRL-2007,GoKr-Nature-2010}.
Here, transient absorption spectroscopy has been used to probe population and coherences within the parent ion~\cite{SaYa-PRA-2011}. 
The high control of the delay between the pump and probe pulses makes it possible to measure even the sub-cycle ionization dynamics and the hole population build-up~\cite{LoGr-ChemPhys-2007,WiGo-Science-2011,PaSy-PRA-2012}.
Also the magnitude of the magnetic quantum number of the hole can be resolved providing information about the hole alignment within an $nl$-subshell~\cite{GoKr-Nature-2010,SaYa-PRA-2011,WiGo-Science-2011}.

In this paper, we show how in photoionization the interference of the direct and indirect ionization pathways results in an unusual ionic state with a highly aligned ionic hole.
%Photoionization studies on Fano resonances studied primarily the photoelectron up to now~\cite{CoMa-PR-1967,ScDo-PRA-1996,LaBe-JPB-1997} to due the technical difficulties to measure the ionic state.
%This has now changed with the advent of attosecond transient absorption spectroscopy.
%In our study, we focus on the parent ion and not on the photoelectron.
Specifically, we consider photoionization of neon with a photon energy that is resonant with the autoionizing $2s^{-1}3p$ state.
We investigate what influence this resonance has on the ionic hole that is eventually formed in the $2p$ shell.
Our calculations are performed using the time-dependent configuration-interaction singles (TDCIS) approach~\cite{GrSa-PRA-2010}.

The correlation-driven autoionization process ($2s^{-1}3p_0 \rightarrow 2p^{-1}_m\,\varepsilon l_m$) produces the same final states as the direct $2p$ photoionization ($2s^22p^6 + \gamma \rightarrow 2p_m^{-1} \varepsilon\, l_m$).
The constructive and destructive interferences of these two pathways lead to the characteristic Fano profile in the photoionization cross section~\cite{CoMa-PR-1967,ScDo-PRA-1996,LaBe-JPB-1997,ArLi-PRL-2010}.
Also the photoelectron angular distribution (characterized by the asymmetry parameter $\beta$) varies strongly across the Fano resonance~\cite{LaBe-JPB-1997}.
In addition, as we demonstrate here, there is also a profound effect on the $2p$ hole, which cannot be deduced from the angular photoelectron distribution.
When tuning the photon energy across the resonance, the hole alignment (the ratio between $2p_0$ and $2p_{\pm 1}$ hole populations) varies dramatically from ratios around 1.6 in the non-resonant case to ratios as large as 19 in the resonant case.

These high ratios, signaling that the hole is dominantly located in the $2p_0$ orbital, are unusual in the XUV regime and are normally only encountered after tunnel ionization with strong-field IR pulses~\cite{LoGr-ChemPhys-2007,IvSp-JMO-2005,Pa-EPJST-2013}.
The maximum hole alignment occurs when the direct $2p$ photoionization pathway interferes most destructively with the indirect $2p$ ionization pathway ($2s^22p^6 + \gamma \rightarrow 2s^{-1}3p_0 \rightarrow 2p^{-1}_m\,\varepsilon l_m$).
The photon energy where this destructive interference is the strongest coincides with the minimum position of the Fano profile.
By tuning the photon energy above or below the $2s$--$3p$ resonance, one controls how these two pathways interfere and, consequently, one controls the $2p$ hole alignment.

%The reason why the $2p_{\pm 1}$ ionization is so much more strongly reduced by this destructive interference than the $2p_0$ ionization can be found by investigating the dominant partial-wave channels $2p^{-1}_m\,\varepsilon d_m$ and the less dominant channel $2p^{-1}_0\,\varepsilon s_0$.
%At the Fano minimum, the $2p^{-1}_m\,\varepsilon d_m$ partial-wave channels are almost completely turned off and only the $2p^{-1}_0\,\varepsilon s_0$ partial-wave channel survives.
%The photoelectron is, therefore, an $s$-wave and the ionic hole is exclusively in the $2p_0$ orbital.

%In the weak-field regime, we discover that a dark state in the continuum can be exploited to control the ionic hole alignment over a wide range.
%Dark states~\cite{MiSi-PRX-2013}, which is normally known from EIT, appear due to a destructive interference between two pathways. 
%In contrast to EIT, only one pathway is light-driven. 
%The other path is purely driven by electron-electron correlations.
%Since electron-electron correlations due not have such a strict rule for the magnetic quantum number $m$ of each individual electron, ionic hole alignments can be generated that are inaccessible to purely light-driven processes using linearly polarized light.

The rest of the manuscript is structured as follows:
Section~\ref{sec2} discusses briefly our TDCIS approach.
In Sec.~\ref{sec3}, we present our results beginning in Sec.~\ref{sec3.fano} with a review of basic aspects of the $2s$--$3p$ Fano resonance in the energy (Sec.~\ref{sec3.fano.spec}) and time domains (Sec.~\ref{sec3.fano.temp}), and explaining in Sec.~\ref{sec3.align} the mechanism of the $2p$ hole alignment when targeting this Fano resonance.
The influence of spin-orbit interaction on the hole alignment is studied in Sec.~\ref{sec3.align.ls}.
Section~\ref{sec4} concludes the discussion.

Atomic units are employed throughout unless otherwise indicated.

\section{Theory}
%------------------------------------------------------------
\label{sec2}

%\subsection{TDCIS}
%\label{sec2.1}

Our implementation of the TDCIS approach to solve the multichannel Schr\"odinger equation has been described in previous publications~\cite{GrSa-PRA-2010,PaGr-PRA-2012}.
We have applied our TDCIS approach to a wide spectrum of processes~\cite{Pa-EPJST-2013}, ranging from attosecond photoionization~\cite{PaSa-PRL-2011} to nonlinear x-ray ionization~\cite{SyPa-PRA-2012} and strong-field tunnel ionization~\cite{WiGo-Science-2011,PaSy-PRA-2012,KaPa-PRA-2013}.

The $N$-body TDCIS wave function reads
\begin{eqnarray}
  \label{eq:tdcis}
  \ket{\Psi(t)}
  &=&
  \alpha_0(t) \, \ket{\Phi_0}
  +
  \sum_{a,i}
    \alpha^a_i(t) \, \ket{\Phi^a_i}
  ,
\end{eqnarray}
where $\ket{\Phi_0}$ is the Hartree-Fock ground state and $\ket{\Phi^a_i}= \hat c^\dagger_a \hat c_i \ket{\Phi_0}$ are singly excited configurations with an electron removed from the initially occupied orbital $i$ and placed in the initially unoccupied orbital $a$.
Since Eq.~\eqref{eq:tdcis} describes all $N$ electrons in the atom, an electron can be removed from any orbital.
This multichannel approach is very helpful in describing ionization processes with XUV and x-ray light where more than one orbital is accessible. 
By limiting the sum over $i$, specific occupied orbitals can be picked to be involved in the dynamics, thereby testing the multichannel character of the overall dynamics.
Inserting Eq.~\eqref{eq:tdcis} into the full time-dependent Schr\"odinger equation, one finds the following equations of motion for the CIS coefficients:

\begin{subequations}
\label{eq:eoms}
\begin{align}
  \label{eq:eoms.1}
  i\, \dot\alpha_0(t)
  =&
  -E(t)\, \sum_{a,i} \sbra{\Phi_0} \hat z \sket{\Phi^a_i}
  \\\nonumber
  \label{eq:eoms.2}
  i\, \dot\alpha^a_i(t)
  =& 
  \sbra{\Phi^a_i} \hat H_0 \sket{\Phi^a_i} \, \alpha^a_i(t) 
  +\!
  \sum_{b,j}
    \sbra{\Phi^a_i} \hat H_1 \sket{\Phi^b_j}
    \alpha^b_j(t)
  \\ &
  -E(t)
  \Big(\!
    \sbra{\Phi^a_i} \hat z \sket{\Phi_0}
    \alpha_0(t)
    +\!
    \sum_{b,j}
      \sbra{\Phi^a_i} \hat z \sket{\Phi^b_j}
      \alpha^b_j(t)
  \!\Big)
  ,
\end{align}
\end{subequations}
where $\hat H_0= \sum_n \left[ \frac{\hat{\bf p}^2_n}{2}  - \frac{Z}{|\hat{\bf r}_n|} + V_\textrm{MF}(\hat{\bf r}_n) - i\eta W(|\hat {\bf r}_n|)\right] - E_\textrm{HF} $ includes all one-particle operators (kinetic energy, attractive nuclear potential, the mean-field potential, $\hat V_\textrm{MF}$, contributing to the standard Fock operator~\cite{SzOs-book}, and the complex absorbing potential, $-i\eta W(\hat r)$, preventing artificial reflection from the boundaries of the numerical grid.
The entire energy spectrum is shifted by the Hartree-Fock energy $E_\textrm{HF}$ such that the Hartree-Fock ground state is at zero energy (for details see Ref.~\cite{RoSa-PRA-2006,GrSa-PRA-2010}).
The nuclear charge is given by $Z$ and the index $n$ runs over all $N$ electrons in the system.
Light-matter interaction for linearly polarized pulses in the electric-dipole approximation is given in the length gauge by $-E(t)\, \hat z$ with $\hat z = \sum_n \hat z_n$~\cite{Pa-EPJST-2013}.
All of the electron-electron interactions that cannot be described by the mean-field potential $\hat V_\textrm{MF}$ are captured by $\hat H_1 = \frac{1}{2}\sum_{n,n'} \frac{1}{|\hat{\bf r}_n - \hat{\bf r}_{n'}|} - \sum_n \hat V_\textrm{MF}(\hat{\bf r}_n)$.
%We introduce a complex absorbing potential that is included in $\hat H_0$~\cite{GrSa-PRA-2010} in order to avoid artificial reflections from the grid boundary.
Introducing a local complex potential has the consequence that the symmetric inner product $\left(\cdot\right|,\left|\cdot\right)$ must be used instead of the hermitian one $\left<\cdot\right|,\left|\cdot\right>$~\cite{RiMe-JPB-1993}.

The second term in Eq.~\eqref{eq:eoms.2}, which describes the electron-electron interaction, is the only term within the TDCIS theory that leads to many-body effects.
Electronic correlation effects, which within TDCIS can only occur between the ionic state (index $i$) and the photoelectron (index $a$), are captured in the interchannel coupling terms ($i \neq j$) where both indices ($a$ and $i$) are changed simultaneously.
It means that the ionic state changes due to the interaction with the excited electron.
Intrachannel interactions do not change the ionic state ($i=j$) and describe the long-range $-1/r$ Coulomb potential for the excited electron.
Intrachannel interaction can be viewed in terms of a one-particle potential and cannot lead to electron-electron correlations. 
The importance of many-body correlation effects~\cite{PaSa-PRL-2011,PaSa-PRL-2013} can be easily tested by either allowing (full TDCIS model) or prohibiting (intrachannel TDCIS model) interchannel interactions which are captured in the $\hat H_1$.

\section{Results}
%------------------------------------------------------------
\label{sec3}

We begin in Sec.~\ref{sec3.fano} with a discussion of the spectral and temporal properties of the $2s$--$3p$ Fano resonance in neon, which we exploit in Sec.~\ref{sec3.align} to control the hole alignment by tuning the XUV pulse across the Fano resonance.

\subsection{$2s$--$3p$ Fano Resonance}
\label{sec3.fano}

\subsubsection{Spectroscopic Features}
\label{sec3.fano.spec}

\begin{figure}[ht!]
  \centering
  \rmpdfinfo
  \includegraphics[width=\figwidth\linewidth]{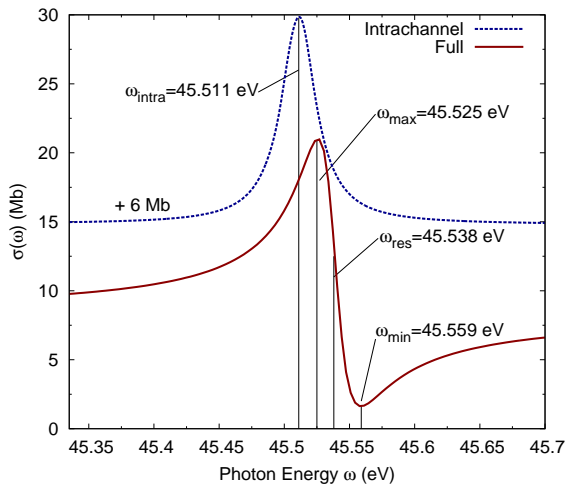}
  \caption{(color online) 
    Photoabsorption cross section of neon in the vicinity of the $2s\rightarrow3p$ resonance for the intrachannel TDCIS model (blue-dashed line) and the full TDCIS  model (red-solid line).  
    Fano profile fits~\cite{LaBe-JPB-1997} give the resonance frequency for the intrachannel TDICS model $\omega_\textrm{intra}$, and the resonance frequency for the full TDCIS model $\omega_\textrm{res}$.
    %the $2s$ hole $T_{2s^{-1}3p}=20.7$~fs. 
    %The minimum position $\omega_\textrm{min}$ and maximum position $\omega_\textrm{max}$ of the full model are visually taken.
    The curve for the intrachannel model is shifted up by +6~Mb for better visualization. 
  }
  \label{fig.cross}
\end{figure}

The photoabsorption cross section, $\sigma(\omega)$, of neon around the $2s$--$3p$ resonance obtained within TDCIS is shown in Fig.~\ref{fig.cross}, both with and without interchannel coupling between the $2s$ and $2p$ shells.
They are both calculated via an autocorrelation function (see Refs.~\cite{Pa-EPJST-2013,KrPa-AJP-2014}).
Strictly speaking, the $2s$--$3p$ resonance has in principle no line width in the intrachanel model since the state $2s^{-1}3p$ cannot autoionize and, therefore, lives forever.
In Fig.~\ref{fig.cross}, this resonance has a finite width that is artificial and has been introduced by hand for better visualization~\footnote{
The autocorrelation function is exponentially damped by hand.
The damping factor is directly related to the width of the Lorentzian.
}.

With the addition of interchannel coupling of the electrons in the full model, the excited $2s^{-1}\,3p$ state autoionizes to a singly charged ionic state $2p^{-1} \, \varepsilon l$.
This indirect ionization of a $2p$ electron ($2s^22p^6+\gamma \rightarrow 2s^{-1}3p \rightarrow 2p^{-1}\,\varepsilon l$) and the direct one-photon ionzation of a $2p$ electron ($2s^22p^6+\gamma \rightarrow 2p^{-1}\,\varepsilon l$) can now interfere, resulting in an asymmetric Fano line shape~\cite{Fa-PhysRev-1961} (see Fig.~\ref{fig.cross}). 
%Recall that no artificial damping was applied to the full model and that the width of the Fano line shape is directly related to the lifetime of the autoionizing state.
We fit both curves (with and without interchannel interactions) to the characteristic Fano profile~\cite{Fa-PhysRev-1961,LaBe-JPB-1997} given by 
\begin{align}
  \label{eq.fano}
  \sigma(\omega)
  =&
  \sigma_a\frac{(q+\epsilon)^2}{1+\epsilon^2}
  +
  \sigma_b
  ,
  \qquad \textrm{with}\quad
  \epsilon = \frac{\omega-\omega_\textrm{res}}{\Gamma/2}
  ,
\end{align}
where $q$ is the Fano parameter describing the asymmetry of the line shape, $\omega_\textrm{res}$ is the resonance frequency of the transition, and $\Gamma$ is the width of the resonance structure. 
These fits give the transition frequencies for both models as well as the transition width and Fano parameter for the full model: $\omega_\textrm{intra}=45.511$~eV, $\omega_\textrm{res}=45.538$~eV, $\Gamma=31.8$~meV, and $q=-1.32$. 
The experimentally obtained value for the resonance position is 45.546~eV, for the line width is 13~meV, and the Fano parameter is $q=-1.58$~\cite{LaBe-JPB-1997,CoMa-PR-1967}.
Since the experimental line width is more than twice as small as our theoretical one, the spectral features presented in Figs.~\ref{fig.holealign_weak}-\ref{fig.align.ls} will be in reality not as broad. Qualitatively, however, this line width discrepancy has no effect on the results and the conclusions.

At frequencies below $\omega_\textrm{res}$, the two ionization pathways constructively interfere and the overall $2p$ ionization is increased.
Above $\omega_\textrm{res}$, the two pathways destructively interfere and the overall $2p$ ionization is suppressed.
The photon energies at the minimum $\omega_\textrm{min}=45.559$~eV and the maximum $\omega_\textrm{max}=45.525$~eV are determined visually.

\subsubsection{Temporal Features}
\label{sec3.fano.temp}
In order to investigate the temporal character of the autoionization process, we resonantly excite neon with a relatively short $2.4$~fs pulse of frequency $\omega_\textrm{res}$ and a peak intensity of $5.6\times 10^{13}$~W/cm$^2$. 
The duration of this pulse is purposely chosen to be much shorter than the lifetime of the $2s$ hole given by $T_{2s^{-1}3p}=1/\Gamma=20.7$~fs, in our calculations. 

\begin{figure}[ht!]
  \centering
  \rmpdfinfo
  \includegraphics[width=\figwidth\linewidth]{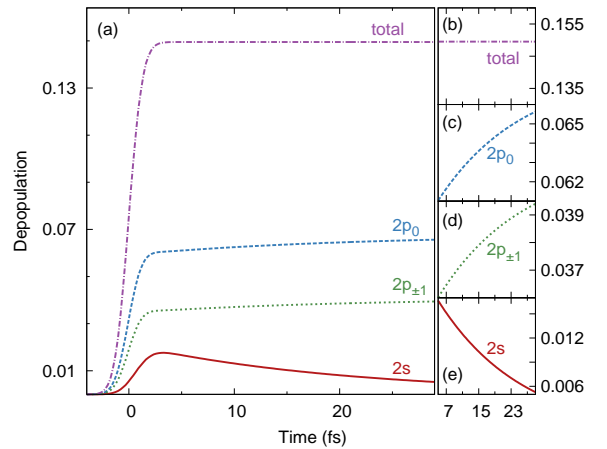}
  \caption{(color online) 
    (a) Hole population of the $2s$ (red-solid line), $2p_0$ (green-dashed line), and $2p_{\pm 1}$ (blue-dashed line) orbitals as well as the ground state depopulation (pink-dashed line) for the full CIS model.
    The pulse has a Gaussian shape and is $2.4$~fs long (FWHM of the intensity) centered around $t=0$, and has the carrier frequency $\omega_\textrm{res}$. 
    Also shown are scaled close-ups of the ground state depopulation (b), and of the hole populations of the different orbitals (c-e).
  } 
  \label{fig.auger}
\end{figure}

The hole population for the $2s$, $2p_0$, and $2p_{\pm 1}$ orbitals as well as the depopulation of the neon ground state are presented in Fig.~\ref{fig.auger}.
Note that for linearly polarized light the sign of the magnetic quantum number $m$ is unimportant and the $+m$ and $-m$ electrons behave exactly in the same way when the initial state is an $M=0$ state as it is the case for closed-shell atoms.
At the end of the pulse, all $2p_m$ depopulations increase while that of the $2s$ decreases. 
The total depopulation, which is the sum of the $2s$ and all $2p_m$ orbitals, remains constant, indicating that the $2p$ and $2s$ hole populations vary equally but oppositely. 
Note that in Fig.~\ref{fig.auger}(b-e) the time scale is changed to visually emphasize these temporal trends.
This is also consistent within TDCIS, where the depopulation of the ground state can no longer change when the pulse is over [see Eq.~\eqref{eq:eoms.1}].
Only the hole rearranges with time from the $2s$ orbital to the $2p$ orbitals.

This hole rearrangement is the resonant Auger decay (or the autoionization process). 
%It is deduced from these curves that the location of the electron transfers from the $2p$ shell to the $2s$ shell. 
The energy released by the hole movement, $26.9$~eV, is sufficient to knock the excited electron residing in the $3p$ shell, which has a binding energy of $2.9$~eV, into the continuum~\footnote{
These energies were calculated using values found in the National Institute of Standards and Technology (NIST) database~\cite{NIST_website}.
}.

\subsection{Hole Alignment}
\label{sec3.align}
As we have seen in Sec.~\ref{sec3.fano.spec}, the indirect ionization pathways via the autoionizing $2s^{-1}3p$ state interferes constructively or destructively with the direct ionization pathway depending on the detuning of the photon energy.
The spectral information (in Fig.~\ref{fig.cross}) does, however, not contain channel-resolved cross sections.
Particularly, it cannot answer the question as to which extent the interference affects all $2p_m$ ionization channels equivalently or whether there is a preferred $m$ ionization channel.
A non-uniform behavior would result in different effective ionization rates for $2p_0$ and $2p_{\pm 1}$ and, consequently, in a modified ratio between $2p_0$ and $2p_{\pm 1}$ hole populations compared to the ratio expected for non-resonant one-photon ionization at similar photon energies.

By studying theoretically and experimentally the angular distribution of the photoelectron~\cite{LaBe-JPB-1997}, a large variation of the asymmetry parameter $\beta$ has been found.
Therefore, we also expect an variation in the ionic hole states.
However, it is not possible to connect directly the angular photoelectron distribution with the ionic hole state.
Theoretical studies~\cite{LaBe-JPB-1997} showed that at $\omega_\textrm{min}$ an asymmetry parameter of $\beta=0$ is expected for the $2s$--$2p$ Fano resonance, meaning the photoelectron is in a pure $s$-wave state.
For this special case, the photoelectron angular distribution can be related to the ionic hole alignment, since an $s$-wave photoelectron can only originate from the $2p_0$ orbital.
Such a connection to the ionic state has, however, not been made in earlier studies.

\begin{figure}[ht!]
  \centering
  \rmpdfinfo
  \includegraphics[width=\figwidth\linewidth]{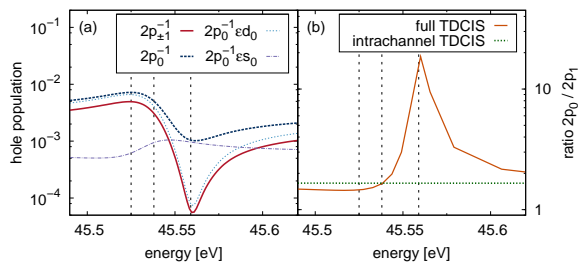}
  \caption{(color online) 
     (a) Hole populations are shown as a function of the photon energy for the $2p_0$ orbital (dashed-dark-blue line), the $2p_{\pm 1}$ orbital (solid-red line) as well as for the two $2p_0$ ionization channels $2p_0^{-1}\varepsilon d_0$ (dotted-light-blue line) and $2p_0^{-1}\varepsilon s_0$ (dashed-dotted-violet line).
     (b) The ratio $2p_0/2p_{\pm 1}$ is shown for the full TDCIS model (solid-orange line) and the intrachannel TDCIS model (dashed-green line).
     A Gaussian pulse with a peak intensity of $3.5\times10^{13}$~W/cm$^2$ and a FWHM-duration of 174~fs has been used.
  }
  \label{fig.holealign_weak}
\end{figure}

In Fig.~\ref{fig.holealign_weak}(a), the $m$-resolved hole populations of the $2p$-shell are shown (thick lines).
Next to the hole populations for $2p_0$ (dashed dark-blue line) and $2p_{\pm 1}$ (solid red line), the two partial-wave channels $2p_0^{-1}\varepsilon s_0$ (dashed-dotted violet line) and $2p_0^{-1}\varepsilon d_0$ (dotted light-blue line) are shown as well.
For the $2p_{\pm 1}$ hole ionization, there exists only one ionization channel where the continuum electron is a $d$-wave (i.e., $2p_{\pm 1}^{-1}\varepsilon d_{\pm 1}$). 

As we can see from Fig.~\ref{fig.holealign_weak}(a) the $2p_m$ populations do vary across the resonance.
Especially the ionization for $2p_{\pm 1}$ is much more suppressed at $\omega_\text{min}$ than for $2p_{0}$.
In the following, we investigate in more details why the ionization of the $2p_m$ orbitals behave so differently by having a closer look at the partial-wave contributions leading to $s$-wave and $d$-wave photoelectrons.

\subsubsection{$d$-wave photoelectron}
The $2p_{\pm 1}$ ionization is much more suppressed than $2p_0$ around $\omega_\textrm{min}$ [see Fig.~\ref{fig.holealign_weak}(a)].
For $2p_{\pm 1}$, the destructive interference is so strong that it leads to a suppression of almost 2 orders of magnitude compared to non-resonant photon energies.
All $2p_m^{-1}\varepsilon d_m$ partial-wave channels show the same degree of suppression.
To be more precise, the ratio between $2p_0^{-1}\varepsilon d_0$ and $2p_{\pm 1}^{-1}\varepsilon d_{\pm 1}$ is exactly $4/3$.
A detailed analysis shows that this ratio between the $m=0$ and $|m|=1$ appears in both, the direct and the indirect, ionization pathways and can be explained by the Wigner-Eckart theorem~\cite{Zare-book}.
%In direct photoionization, the transition rate of $2p_m \rightarrow \varepsilon d_m$ is proportional to $[C^{2,m}_{1,m;1,0}]^2$ according to the Wigner-Eckard theorem, where $C^{l_3,m_3}_{l_1,m_1;l_2,m_2}=\braket{l_1,m_1;l_2,m_2}{l_3,m_3}$ is the Clebsch-Gordan coefficient.
%Hence, one finds the ratio $2p_0/2p_1 = 2p_0/(2p_{-1}+2p_{+1}) = [C^{2,0}_{1,0;1,0}/C^{2,+1}_{1,+1;1,0}]^2/2=2/3$.
%For the autoionization $2s^{-1}\,3p_0 \rightarrow 2p_m^{-1} \varepsilon d_m$ the same ratio will be found between the $m=0$ and the $m=\pm1$ channels.
Consequently, the behavior of constructive and destructive interference is exactly the same for all $d$-wave channels, $2p_m^{-1} \varepsilon d_m$.

\subsubsection{$s$-wave photoelectron}
To generate $2p_0$ holes there exists another ionization channel leading to an $s$-wave photoelectron, i.e., $2p_0^{-1} \varepsilon s_0$.
The behavior of this partial-wave channel is different than the behavior of the $2p_m^{-1} \varepsilon d_m$ partial-wave channels [see Fig.~\ref{fig.holealign_weak}(a)].
For $2p_0^{-1} \varepsilon s_0$, the destructive interference happens at $\omega_\textrm{max}$ and constructive interference occurs at $\omega_\textrm{min}$.

The overall trend is dominated by $2p_m^{-1} \varepsilon d_m$, since the probability of ejecting an electron from a $p$-orbital into an $s$-continuum is generally much smaller than ejecting the electron into a $d$-continuum~\cite{FaCo-RMP-1968}.
Only around $\omega_\textrm{min}$, where the ionization into a $d$-continuum is strongly suppressed, the situation changes and ionization into the $s$-continuum becomes the dominant ionization channel (corresponding to an asymmetry parameter of $\beta=0$).
The relative enhancement of the $2p_0^{-1} \varepsilon s_0$ partial-wave channel results in a ten times smaller overall suppression for $2p_0$ ionization than for $2p_{\pm 1}$ ionization [see Fig.~\ref{fig.holealign_weak}].

\subsubsection{The ratio of $2p_m$ hole populations}
In Fig.~\ref{fig.holealign_weak}(b), the hole population ratio $2p_0/2p_{\pm 1}$ is shown as a function of the photon energy for the full TDCIS model (orange-solid line) and the intrachannel TDCIS model (green-dashed line).
This ratio is a direct measure of hole alignment, where 1 stands for an isotropic hole distribution, $\infty$ for perfect hole alignment along the polarization direction, and 0 for perfect hole antialignment in the plane perpendicular to the polarization direction.

Strong variations of the hole alignment across the Fano resonance are found resulting in ratios that vary by more than one order of magnitude (between 1.6 and 18).
A ratio of 18 means the $2p$ hole is primarily located in the $2p_0$ orbital, and only a 10\% chance exists to find the hole in either the $2p_{+1}$ or $2p_{-1}$ orbital.
Such strong hole alignment is normally only encountered in the strong field regime where tunnel ionization almost exclusively ionizes the outermost $p_0$ orbital (when using linearly polarized light)~\cite{PaGr-PRA-2012,IvSp-JMO-2005}.

In the off-resonance limit, the intrachannel TDCIS model and the full TDCIS model approach the same value for the $2p_0/2p_{\pm 1}$ ratio (1.6).
Such values are very common in the XUV and x-ray regimes where an almost isotropic distribution of the hole is found with a slight preference for the polarization direction (i.e., $m=0$).

The maximum hole alignment is reached when the photon energy is $\omega_\textrm{min}$, located at the minimum of the Fano resonance, which is exactly the energy where the suppression of the dominant ionization channels (leading to $2p^{-1}_m\,\varepsilon d_m$) is most pronounced, and only $s$-wave photoelectrons are formed which leave a $2p_0$ hole behind.

\subsection{Spin-orbit coupling}
\label{sec3.align.ls}
Up to now, we have ignored that the $2p$ shell is actually split due to spin-orbit coupling into two subshells $2p_j$ with $j=1/2$ and $j=3/2$.
As a result, the hole alignment has to be defined with respect to $m_j$ and not $m_l$.
In particular, the $2p^{m_j}_{3/2}$ hole populations for $m_j=\pm 1/2$ and $m_j=\pm 3/2$ have to be compared.
Here, $m_j$ refers to the projection of the total angular momentum $j$ along the XUV polarization axis.
In our TDCIS approach, we consider only the spin-orbit interaction within the ion, where it is the strongest, and we neglect it for the photoelectron (see Ref.~\cite{PaSy-PRA-2012} for details).

\begin{figure}[ht!]
  \centering
  \rmpdfinfo
  \includegraphics[width=\figwidth\linewidth]{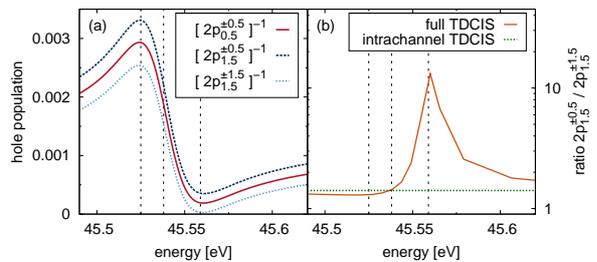}
  \caption{(color online) 
     (a) Hole population is shown as a function of the photon energy for the $2p^{\pm0.5}_{0.5}$ orbital (solid-red line), the $2p^{\pm0.5}_{1.5}$ orbital (dashed-blue line), and the $2p^{\pm1.5}_{1.5}$ orbital (dotted-light-blue line).
     (b) The ratio $2p^{\pm1.5}_{1.5}/2p^{\pm 0.5}_{1.5}$ is shown for the full TDCIS model (solid-brown line) and the intrachannel TDCIS model (dashed-green line).
     The same pulse parameters as in Fig.~\ref{fig.holealign_weak} has been used.
  }
  \label{fig.align.ls}
\end{figure}

Figure~\ref{fig.align.ls} (a) shows the hole populations of $2p^{\pm0.5}_{0.5}$, $2p^{\pm0.5}_{1.5}$, and $2p^{\pm1.5}_{1.5}$, and (b) shows the ratio between $2p^{\pm0.5}_{1.5}$ and $2p^{\pm1.5}_{1.5}$ defining the hole alignment.
Figure~\ref{fig.align.ls} shows the same trends as Fig.~\ref{fig.holealign_weak}. 
The mixing of $2p_0$ and $2p_{\pm 1}$ orbitals in the spin-orbit case reduces the maximum hole alignment within the $2p_{3/2}$-shell by $\sim\!2/3$ in comparison to the non-spin-orbit case~\footnote{The exact transformation from the hole alignment ratio without spin-orbit coupling, $r_\textrm{nols}$, to the ratio when spin-orbit coupling is included, $r_\textrm{ls}$, reads: $r_\textrm{ls} = 1/3 + 2/3\, r_\textrm{nols}$. For large hole alignments, the constant can be ignored.}, 
which results in a maximum alignment ratio of $\sim\!13$ instead of 18.

The reduction factor of 2/3 can be easily explained when expressing the spin-orbit-split orbitals in terms of the non-spin-orbit-split orbitals.
Specifically, the transformation between the spin-orbit-split (coupled basis) and non-spin-orbit-split (uncoupled basis) orbitals reads:

\begin{subequations}
\label{eq:pop.ls}
\begin{align}
  \label{eq:pop.ls_0.5-0.5}
  \ket{2p^{\pm 0.5}_{0.5}}
  =&  
  \pm\sqrt{\frac{2}{3}} \ket{2p_{\pm 1,\mp \half}\big.}
  \mp
  \sqrt{\frac{1}{3}} \ket{2p_{0,\pm \half}\big.}
  \\
  \label{eq:pop.ls_0.5-1.5}
  \ket{2p^{\pm 0.5}_{1.5}}
  =&  
  +\sqrt{\frac{1}{3}} \ket{2p_{\pm 1,\mp \half}\big.}
  +
  \sqrt{\frac{2}{3}} \ket{2p_{0,\pm \half}\big.}
  \\
  \label{eq:pop.ls_1.5-1.5}
  \ket{2p^{\pm 1.5}_{1.5}}
  =&
  \ket{2p_{\pm 1,\pm \half}\big.}
\end{align}
\end{subequations}
where $\ket{2p_{m,\sigma}}$ refers to the spatial $2p_m$ orbital with the spin projection $\sigma$.
Note that in Sec.~\ref{sec3.align} we focused only the spatial part of the orbitals because the spin-up and spin-down components behave exactly the same~\footnote{To be more precise, the non-spin-orbit-split hole populations $2p_m$ are the sums of the corresponding spin-up and spin-down hole populations.}.
The spin-orbit interaction is treated here in degenerate perturbation theory (see Ref.~\cite{RoSa-PRA-2009,PaSy-PRA-2012}) where only the impact on the angular momentum is considered.
The radial part is unaffected by the spin-orbit interaction which leads to errors of few per cent~\cite{PaSa-JPB-submitted}.

By using Eqs.~(\ref{eq:pop.ls_0.5-0.5}--\ref{eq:pop.ls_1.5-1.5}), all populations shown in Fig.~\ref{fig.align.ls}(a) can be written in terms of the non-spin-orbit-split populations shown in Fig.~\ref{fig.holealign_weak}(a), and, consequently, also the alignment ratio in the case of spin-orbit splitting can be expressed in terms of the ratio without spin-orbit splitting as done earlier.

\section{Conclusion}
\label{sec4}

We have shown that resonant excitation of the autoionizing $2s^{-1}3p$ state leads to a second ionization pathway that can interfere with the direct $2p$ photoionization pathway and strongly influences the state of the parent ion.
This interference is well known as the origin of the characteristic Fano profile.
Also the asymmetry parameter $\beta$ measuring the angular distribution of the photoelectron varies strongly across the Fano resonance but a direct relation to the hole alignment cannot be made.

We showed that this interference has destructive character at $\omega_\textrm{min}$ and creates a dark-state in the photoelectron continuum. 
As a result, the $2p^{-1}\varepsilon d$ ionization channel is strongly suppressed, and the photoelectron is emitted as a pure $s$-wave.
Consequently, the only orbital that is ionized is the $2p_0$ orbital.
The imbalance of ionizing $2p_0$ and $2p_{\pm1}$ orbitals leads to a large hole alignment along the XUV polarization direction.

The ratio between the populations of $2p_0$ and $2p_{\pm1}$ goes as high as 19---localizing the hole in the $2p_0$ orbital---and is significantly different than the off-resonant value ($1.6$), which possesses only a slight hole alignment.
Strong hole alignments are usually only encountered after tunnel ionization with strong-field IR pulses, where the Keldysh parameter is well below 1~\cite{Pa-EPJST-2013}.
Here, we used XUV pulses and we are in the perturbative one-photon regime, where the Keldysh parameter is well above 1 and large anisotropies in the hole states are not expected. 

When disabling interchannel coupling effects, i..e, disabling the correlation-driven autoionization mechanism of the excited $2s^{-1}3p$ state, no interference of the ionization pathways occurs and no hole alignment modulation appears when tuning across the $2s$--$3p$ resonance.
Including spin-orbit interaction within the ion does not change the picture.
Only the strong hole alignment within the $2p_{3/2}$-shell is reduced by a factor $2/3$, which still results in a strong hole alignment with ratios up to $13\!:\!1$ between $2p_{3/2}^{\pm1/2}$ and $2p_{3/2}^{\pm3/2}$ hole populations.

Controlling the hole alignment via the $2s$--$3p$ Fano resonance serves as an example of how correlation effects can be explicitly targeted and exploited to create new and exotic electronic states in atoms and molecules.
Similiarly other Fano resonances can be used where the strength of the resonance determines how strongly the hole alignment can be tuned.
Furthermore, with a second pulse the Fano resonance could be modified within attoseconds~\cite{OtKa-Science-2013} to gain an even larger control of the electronic motion.
Also the extension to high-intensity pulses is interesting, which can be realized with currently available seeded free-electron lasers like FERMI~\cite{MuNi-NatPhoton-2012} or sFLASH~\cite{DeDr-PRL-2013}.
First preliminary results we have obtained suggest that completely different ionization behavior occurs when a Fano resonance is driven by a high-intensity pulse.

\acknowledgments
E. H.-J. would like to thank the DAAD RISE program and her mother Christa Heinrich-Josties for financial support.
This work has been supported by the Deutsche Forschungsgemeinschaft (DFG) under grant No. SFB 925/A5.

%% use bibtex
\bibliographystyle{apsrev4-1-etal}
\bibliography{amo,books,notes}

%% include reference for manual modification 
%\input{citations}

\end{document}